\documentclass[twocolumn,twoside,slac_two]{revtex4}
\usepackage{graphicx}
\usepackage{fancyhdr}
\usepackage{graphics}
\usepackage{epstopdf}
\usepackage{textpos}
\usepackage{atlasphysics}
\usepackage{xspace}
\newcommand{\sqR}   {\tilde{q}^{}_{R}}
\newcommand{\sqL}   {\tilde{q}^{}_{L}}

\newcommand{\bone}   {\tilde{b}^{}_{1}}
\newcommand{\tone}   {\tilde{t}^{}_{1}}
\newcommand{\neut}  {\tilde{\chi}^{0}_{1}}  
 
\def\etmiss{\ensuremath{E_{\mathrm{T}}^{\mathrm{miss}}}\xspace}

\def\dphimin{\ensuremath{\Delta\phi_{min}}}

\def\meff{\ensuremath{\mathrm{m}_{\mathrm{eff}}}\xspace}

\newcommand{\gl}   {\tilde{g}}
\newcommand{\sq}   {\tilde{q}}

\renewcommand{\ttbar} {\ensuremath{t\bar{t}}\xspace}

\renewcommand{\pt} {\ensuremath{p_\mathrm{T}}\xspace}

\newcommand{\lumi}{0.83}

\begin{document}

\title{\centering Search for New Physics at $\sqrt{s}$ = 7 TeV in Hadronic Final States with Missing Transverse Energy and Heavy Flavor}


\author{
\centering
\begin{center}
B. Butler, on behalf of the ATLAS Collaboration
\end{center}}
\affiliation{\centering SLAC National Accelerator Laboratory, California, 94025, USA}
\begin{abstract}
A search for supersymmetric particles in events with large missing transverse 
momentum, heavy flavor jet candidates and no leptons ($e$,$\mu$) in $\sqrt{s} = 7$~TeV 
proton-proton collisions is presented. In a data sample corresponding to an integrated luminosity of 0.83 ~fb$^{-1}$ recorded by the 
ATLAS experiment at the Large Hadron Collider, no significant excess is observed with respect to the prediction for Standard Model processes.
Model-independent production cross section upper limits are provided in the context of simplified 
models as well as conventional limits.
\end{abstract}

\maketitle
\thispagestyle{fancy}


\section{Introduction}
Supersymmetry (SUSY)~\cite{Golfand:1971iw} is one of the most compelling theories 
to describe physics beyond the Standard Model (SM). In the framework of a generic $R$-parity conserving minimal 
supersymmetric extension of the SM, the MSSM~\cite{Martin:1997ns}, SUSY particles are produced in pairs and 
the lightest supersymmetric particle (LSP) is stable.
The colored superpartners of quarks and gluons, the squarks ($\sq$) and 
gluinos ($\gl$), are expected to be copiously produced via the strong 
interaction at the Large Hadron Collider (LHC).
The partners of the right-handed and left-handed quarks, $\sqR$ and $\sqL$, 
can mix to form two mass eigenstates. These mixing effects are proportional to 
the corresponding fermion masses and therefore become important for the third 
generation. In particular, large mixing can yield sbottom ($\bone$) and stop ($\tone$) 
mass eigenstates that are significantly lighter than other squarks. 
Consequently, $\bone$ and $\tone$ could be produced with large cross sections at 
the LHC. Direct $\sq$ pair production or $\gl\gl$ production with subsequent $\gl \rightarrow \bone b$ or  $\gl \to \tone t$ decays 
result in complex final states consisting of 
missing transverse momentum (its magnitude is referred to as 
$\etmiss$ in the following) and several jets.  Among these jets, $b$-quark jets ($b$-jets) 
are expected. 

SUSY is searched for in final states involving $\etmiss$, 
energetic jets, of which at least one must be identified as a $b$-jet and no 
isolated leptons ($e$ or $\mu$). The search is based on $pp$ collision 
data at a center-of-mass energy of 7~TeV recorded by the ATLAS experiment~\cite{DetectorPaper:2008} at the 
LHC in 2011. The total data set used corresponds to an integrated luminosity of 
\lumi~$\ifb$.

Two phenomenological 
MSSM scenarios are considered where the first and second generation squark masses are set 
above 2 TeV. In the first scenario, the $\bone$ is the lightest squark,
$m^{}_{\gl}>m^{}_{\bone}>m^{}_{\neut}$, and the branching ratio 
for $\gl \rightarrow \bone b$ decays is 100\%. Sbottoms are produced 
via gluino-mediated processes or via direct pair production and they 
are assumed to decay exclusively via $\bone \rightarrow b\neut$, where 
$m^{}_{\neut}$ is fixed at 60~GeV. The interpretation of the results is presented as a function of the gluino and light sbottom masses. 

The second MSSM-like scenario is defined in the context of the general 
simplified models~\cite{Alves:2011wf}: all squarks including $\bone$ are heavy, 
gluino-pair production is the only kinematically allowed process and 
gluinos decay (off-shell) into~$b\bar{b}\neut$ final states. Here the results are interpreted in a  ($m_{\gl},m_{\neut}$) plane. 
These results are generalized to any new physics process where  
gluino-like particles decay into $b\bar{b}$ and a weakly interacting 
massive particle.

\section{Simulation and Event Selection}\label{sec:SimEvSamp}
Simulated event samples are used to determine the detector acceptance, the 
reconstruction efficiencies and the expected event yields. Details can be found in Ref.~\cite{ATLAS-CONF-2011-098}. For the background, the following Standard Model processes are considered: 
\begin{itemize} 
\item $\ttbar$ and single top production
\item $W (\rightarrow \ell \nu)$+jets, $Z/\gamma^*(\rightarrow \ell^+ \ell^-)$+jets  
(where $\ell=e,\ \mu,\ \tau$) and 
$Z(\rightarrow \nu \bar{\nu})$+jets  
production
\item Di-boson ($WW$, $WZ$ and $ZZ$) production, which is found to be negligible. 
\end {itemize} 

\noindent For the QCD background, no reliable prediction can be obtained from leading-order 
Monte Carlo simulation. A data-driven method is used as discussed in Section~\ref{sec:wbkg}.


Events are selected at the trigger level by 
requiring one jet with high $\pt$ and large missing transverse momentum. The selection is fully efficient 
for events containing at least one jet with $\pt >130$~GeV and $\etmiss>130$~GeV. 
Only jets with $\pt>20$~GeV and within $|\eta|<2.8$ are retained.
Candidates for $b$-jets are identified among jets with $\pt>50$~GeV using 
a secondary vertex with a tagging efficiency of 50\% (1\%) for $b$-jets (light flavor or gluon jets) in $\ttbar$ events in simulation. 
A lepton veto is applied to electrons (muons) with $\pt>20$ GeV (10 GeV) and $|\eta|<2.47 (2.4)$. 

The effective mass, $\meff$, is defined as the scalar sum of $\etmiss$ and the 
transverse momenta of the three leading jets. Events are required to 
have $\etmiss / \meff>0.25$. In addition, the smallest azimuthal separation between the 
$\etmiss$ direction and the three leading jets, $\dphimin$, is required to be larger 
than 0.4 to reduces the amount of QCD  background. Full selection details can be found in Ref.~\cite{ATLAS-CONF-2011-098}.

\section{Signal Region Optimization}

Phenomenological models describe well-motivated, SM-like production and decay processes, and the kinematics are determined by a small number of parameters (masses). Both SUSY scenarios considered in this analysis result in 4 $b$-jet + $\etmiss$ final state signatures. The simplified model kinematics are determined to first order by the mass difference between the gluino and the neutralino. This simplicity motivates the choice of this model for signal region optimization studies, with cross-checks to ensure the results are relevant to the gluino-squark on-shell cascade case.

Signal region optimization consists of an $n$-dimensional significance maximization procedure to produce a set of optimal selections which were then iteratively reduced in number while ensuring broad-based sensitivity was retained. Four signal regions were chosen to represent good compromises among mass plane coverage, sensitivity, and practical concerns such as background control regions. They are characterized by the minimum number of 
$b$-jets required (1,2) and by the \meff threshold (500, 700 GeV).

\section{Background Estimation and Systematic Uncertainties}\label{sec:wbkg}\label{syst}
Events from \ttbar production represent the largest background component in all four signal regions. The Monte Carlo 
prediction is validated by a data-driven estimate which relies on 1-lepton control regions with similar kinematic selections to those of the signal regions~\cite{ATLAS-CONF-2011-098}. The normalization 
determined in these control regions (corrected for non-$\ttbar$ contamination) is then transferred to the signal regions. 



The Monte Carlo estimation of the $W/Z$ background is 
validated with a combined fit of $\ttbar$ and $W/Z$ background components to the distribution of the number of $b$-tagged jets in a 0-lepton control region defined by reversing the $\meff$ cut. 
The total systematic uncertainty on the Monte Carlo prediction 
is estimated to be between $\pm$30\% and $\pm$35\% depending on the final selection, and is dominated by the jet energy scale, theoretical, and b-tagging efficiency uncertainties \cite{ATLAS-CONF-2011-098}. 

The remaining QCD background in the signal regions is estimated with a data driven procedure. 
The technique~\cite{daCosta:2011qk,ATLAS-CONF-2011-086} used is to smear the momentum of jets in clean data 
events with low $\etmiss$ to generate "pseudoevents" with potentially large $\etmiss$
values.
The method was validated by comparing data and pseudoevents distributions in QCD enriched control 
regions obtained by reversing the cut 
on $\dphimin$. 
The uncertainty of 50\% is dominated by the dependency 
of the smearing function on the flavor composition of the low \etmiss sample \cite{ATLAS-CONF-2011-098}.




\section{Results}
Good agreement between data and Monte Carlo prediction is observed in the $\meff$ distribution (Figure~\ref{fig:datamc3jSRs}) for the signal regions with two $b$-tags. A similar level of agreement is observed for signal regions with one $b$-tag.
\begin{figure}
  \centering
\includegraphics[width=1.0\columnwidth]{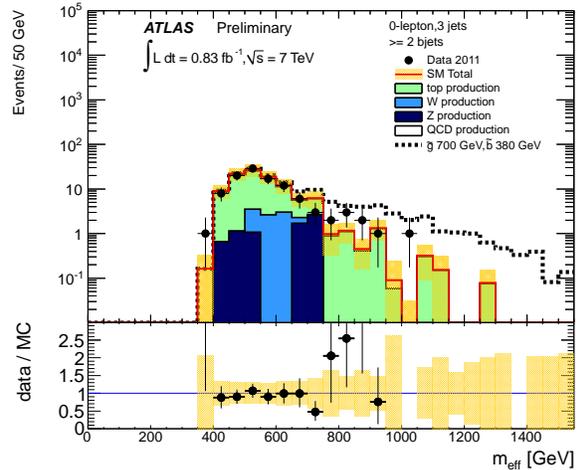}
  \caption{The $\meff$ distribution for data and the expected SM processes in the 2 $b$-tags signal regions.  The yellow band shows 
the full systematic uncertainty on the SM expectation. For illustration, the distributions 
of one reference SUSY signal are superimposed.}\label{fig:datamc3jSRs}
\end{figure}

The observed and predicted event yields in the four signal regions are given in Table~\ref{tab:3jresults}. The SM predictions agree with the observed number of events in all four signal regions. 
 
\begin{table}[htp]
 \centering
 \begin{tabular}{c|c|c|c|c|c}
   \hline
   \hline
   Sig. Reg. & Data & Top & W/Z &  QCD & Total \\
   \hline
   3JA (1,500) & 361 & $221^{+82}_{-68}$ & $121 \pm 61$ & $15 \pm 7$ & $356^{+103}_{-92}$\\
   3JB (1,700) & 63 & $37^{+15}_{-12}$ & $31 \pm 19$  & $1.9 \pm 0.9$ & $70^{+24}_{-22}$\\
   3JC (2,500) & 76 & $55^{+25}_{-22}$ & $20 \pm 12$ & $3.6 \pm 1.8$ & $79^{+28}_{-25}$\\
   3JD (2,700) & 12 & $7.8^{+3.5}_{-2.9}$ & $5 \pm 4$  & $0.5 \pm 0.3$& $13.0^{+5.6}_{-5.2}$\\
   \hline
   \hline
   \end{tabular}
   \caption{Summary observed and expected event yields in the four signal regions. The selections 
   differentiating the regions are given after the region name in the format (n $b$-tag(s), $\meff$ cut in GeV). }
\label{tab:3jresults}
\end{table}

95\% C.L. exclusion imits are derived using the $CL_s$~\cite{Read:2002hq} method, while the power constrained limit 
(PCL)~\cite{Cowan:PCL} method is used for comparison with previous ATLAS results. For each scenario, the signal region resulting in the best expected exclusion limit is used.
In Figure~\ref{fig:sb_gl_obs} the observed and expected exclusion regions 
are shown in the ($m^{}_{\gl},m^{}_{\bone}$) plane for the hypothesis that the lightest 
squark $\bone$ is produced via gluino-mediated or direct pair production and 
decays exclusively via $\bone \rightarrow b\neut$. Gluino masses below 720~GeV are excluded for sbottom masses up to 600~GeV.
This search extends the previous ATLAS exclusion limit by about 130~GeV \cite{Aad2011398}.  

Results are also interpreted in the context of simplified models. In this case, all the squarks are heavier than the gluino, which decays exclusively into three-body final states ($b\bar{b}\neut$) via an off-shell sbottom.  
The exclusion limits obtained on the ($m^{}_{\gl},m^{}_{\neut}$) plane are shown in Figure~\ref{fig:GbbMaxAll}, as well as $\sigma\times BR$ upper limits.  
Neutralino masses below 200-250~GeV are 
excluded for gluino masses in the range 200-660~GeV, if $\Delta M (\gl-\neut)>$100 GeV. 

\begin{figure}[tb]
  \begin{center}
      \includegraphics[width=1.0\columnwidth]{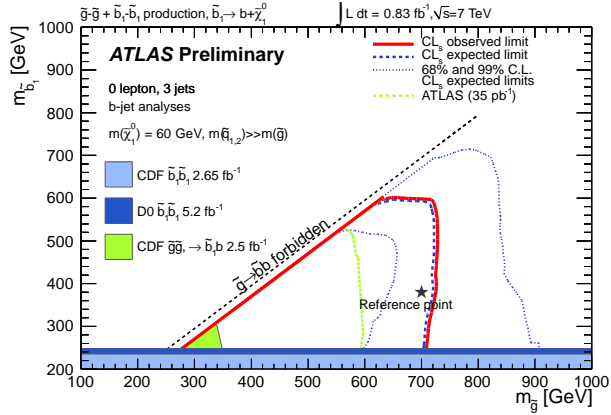}
    \end{center}
  \caption{Observed and expected 95\% C.L. exclusion limits in the ($m^{}_{\gl},m^{}_{\bone}$) plane. 
  The neutralino mass is set to  60~GeV. 
  The result is compared to previous ATLAS and CDF results. Exclusion limits from the CDF and D0 experiments on direct sbottom pair production are 
  also shown. }
  \label{fig:sb_gl_obs}
\end{figure}
\begin{figure}[h!]
 \begin{center}
     \includegraphics[width=1.0\columnwidth]{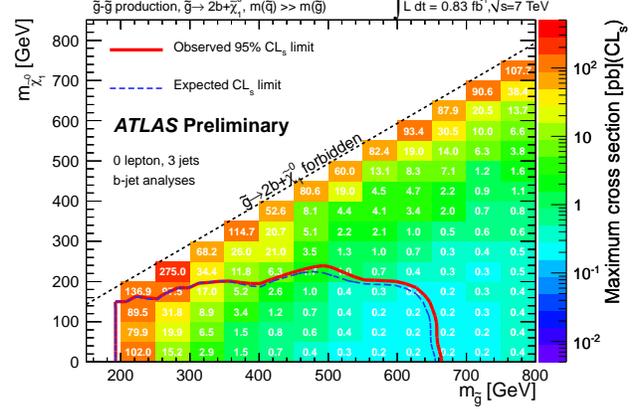} 
   \end{center}
 \caption{95\% C.L. upper cross section limits in pb and observed and expected limit contours  
  in the ($m^{}_{\gl},m^{}_{\neut}$) plane for gluino masses above 200 GeV. }
  \label{fig:GbbMaxAll}
\end{figure}

\section{Conclusions}
A search for SUSY in final states with missing transverse 
momentum, $b$-jets and no isolated leptons in pp collisions 
at 7~TeV is presented. No excess above the expectation from SM processes is found. The 
results are used to exclude parameter regions in various $R$-parity conserving SUSY models. 

\begin{acknowledgments}
This proceeding is adapted from an ATLAS conference note \cite{ATLAS-CONF-2011-098}, and as such, includes text written by several members of the ATLAS collaboration.
\end{acknowledgments}

\bigskip 




\end{document}

%